\documentclass[doublecol]{epl2}
\pdfoutput=1

\usepackage{amssymb}
\usepackage{amsmath,amsthm}
\usepackage{color}

\title{Cooperative Motion of Active Brownian Spheres in Three-Dimensional Dense Suspensions}
\shorttitle{Cooperative Motion of Active Brownian Spheres}

\author{Adam Wysocki \and Roland G. Winkler \and Gerhard Gompper}
\shortauthor{Adam Wysocki \etal}

\institute{Theoretical Soft Matter and Biophysics,
Institute for Advanced Simulation and Institute of Complex Systems,
Forschungszentrum J\"ulich, 52425 J\"ulich, Germany}

\abstract{The structural and dynamical properties of suspensions of
self-propelled Brownian particles
of spherical shape are investigated in three spatial dimensions.
Our simulations reveal a phase separation into a
dilute and a dense phase, above a certain density and strength of self-propulsion.
The packing fraction of the dense phase approaches random close packing
at high activity, yet the system remains fluid. 
Although no alignment mechanism exists in this model, we find
long-lived cooperative motion of the particles in the dense regime.
This behavior is probably due to an interface-induced sorting process.
Spatial displacement correlation functions are nearly scale-free
for systems with densities close to or above the glass transition density
of passive systems.}

\begin{document}

\maketitle

\section{Introduction}
Assemblies of intrinsically active objects, sometimes called {\it
living fluids}, represent an exceptional class of non-equilibrium
systems. Examples range from the macroscopic scale of human crowds
to the microscopic scale of cells and motile microorganisms such
as bacteria\cite{vicsek2012pr,angelini2011pnas}. A generic
phenomenon of dense living fluids is the emergence of
self-organized large-scale dynamical patterns like vortices,
swarms, networks, or self-sustained turbulence
\cite{kearns2010nrm,wensink2012pnas}. This intriguing dynamical
behavior is a consequence of the complex interplay of
self-propulsion, internal or external noise, and many-body
interactions.

The understanding of the collective behavior requires the
characterization of the underlying physical interaction
mechanisms. Experiments and simulations indicate that alignment
induced by particle interactions, e.g., inelastic collisions
between elongated objects
\cite{peruani2006pre,yang2010pre,wensink2012pnas} or hydrodynamic
interactions \cite{lauga2009rpp}, lead to clustering and
collective motion. Studies of rodlike self-propelled particles in
two dimensions (2D) revealed mobile clusters and a variety of
other dynamical phases
\cite{peruani2006pre,yang2010pre,wensink2012pnas}. In contrast,
systems of disc-like particles in 2D, which lack an alignment
mechanism, do not form any ordered moving states. Yet, they
exhibit an activity-induced phase transition. At lower densities
small, transient clusters appear
\cite{fily2012prl,theurkauff2012prl,palacci2013science,buttinoni2013prl}.
Above a critical density, the system separates into a dense and a
dilute phase \cite{fily2012prl,redner2013prl,bialke2013epl,speck2013arxiv}. 
The dense phase exhibits a crystalline order
with a slow dynamics, mainly due to movements of dislocations
\cite{redner2013prl,palacci2013science}.
Less attention has been paid to three-dimensional (3D) suspensions
of spherical (colloid-like) active particles (with the exception of a 
very recent numerical \cite{stenhammar2014sm} and theoretical work \cite{cates2013epl}). 
In such systems, the question naturally arises, whether a
collective behavior emerges despite the particles lack any
aligning interactions. In this letter, we present simulation
results of the emergent structure and dynamics of motile spherical
particles in 3D---as a generic model of non-aligning self-driven
particles. We find that an isotropic system phase separates into a
dilute, gas-like phase, and a dense, liquid-like phase at
sufficiently large activity. Inside the dense phase, the particles
exhibit collective motion---in the form of jets and swirls---even
at densities very close to random close packing (RCP). Although a
global polar order is absent in the system, long-lived cooperative
motion with nearly scale-free correlations emerge.

\begin{figure*}[t]
\centering
\includegraphics[width=1.6\columnwidth]{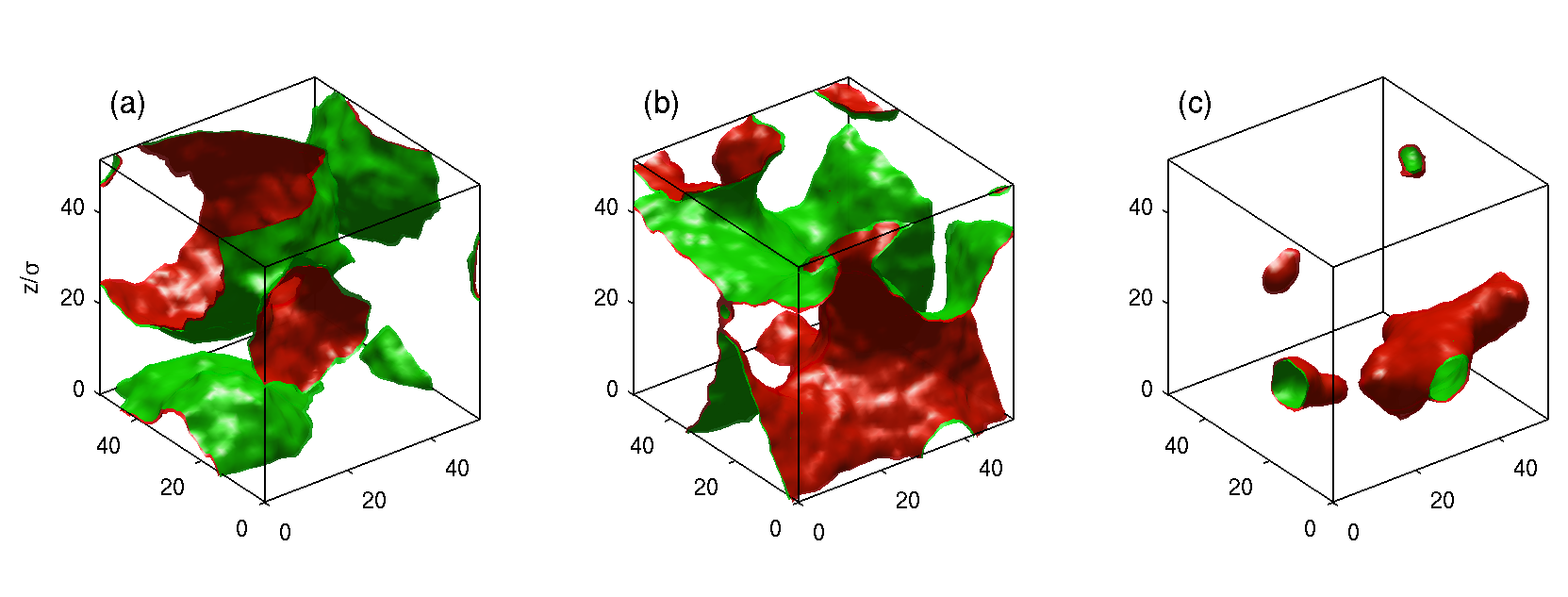}
\caption{\label{f:isodensity}(color online) Snapshots of the gas-liquid interface of a 
system (a,b) just above the clustering transition ($\phi=0.3375$ and $\phi=0.4$) and (c) at a 
high density ($\phi=0.6$). The red isodensity surfaces is drawn at $\phi_{loc}=0.54$, 
the green at $\phi_{loc}=0.5$. See also movies {\tt movie1.avi} and {\tt movie2.avi}.}
\end{figure*}

\section{Model}
We model an active particle as a hard-sphere-like particle
which is propelled  with constant velocity $V_{0}$ along its
orientation vector $\mathbf{e}$ in 3D
\cite{bialke2012prl,fily2012prl,redner2013prl}.
Its translational motion is described by
\begin{equation}
\dot{\mathbf{r}}=V_{0}\mathbf{e}+\mathbf{F}/\gamma_{t}+\boldsymbol{\eta},
\label{e:BDtrans}
\end{equation}
where $\dot {\mathbf{r}}$ is the velocity, $\mathbf{F}$ the  total
interparticle force, and $\boldsymbol{\eta}$ a Gaussian
white-noise random velocity, with
$\langle\boldsymbol{\eta}\rangle=0$ and
$\langle\boldsymbol{\eta}_i(t)\boldsymbol{\eta}_j(t')\rangle=
2D_{t}\mathbf{1}\delta_{ij}\delta(t-t')$. The translational
diffusion coefficient $D_t$ is related to the frication
coefficient $\gamma_t$ via $D_t =k_BT/\gamma_t$. The
force follows from a steep (shifted) Yukawa-like potential
$V(r)/k_BT=5 e^{-\kappa(r-\sigma)}/[\kappa(r-\sigma)]$ for
$r>\sigma$, $V(r)= \infty$ for $r\le\sigma$, and $V(r)=0$ for
$r>1.083\sigma$, with particle diameter $\sigma$ and
$\kappa\sigma=60$.
The orientation $\mathbf{e}$ performs a random walk according to
$\dot{\mathbf{e}}=\boldsymbol{\xi}\times\mathbf{e}$, where $\boldsymbol{\xi}$
is a Gaussian white-noise vector, with $\langle\boldsymbol{\xi}\rangle=0$ and
$\langle\boldsymbol{\xi}_i(t)\boldsymbol{\xi}_j(t')\rangle=2D_{r}\mathbf{1}\delta_{ij}\delta(t-t')$.
The translational and rotational diffusion coefficients, $D_{t}$ and $D_{r}$,
are related via $D_{t}=D_{r}\sigma^2/3$. The importance of noise
is measured by the P\'eclet number $Pe=V_0/(\sigma D_r)$;
we study systems in the range of $9 \leq Pe \leq
272$. Up to $N=1.5\times 10^{5}$ particles are simulated in a
cubic box of length $L$. The density is measured in terms of the
global packing fraction $\phi=\pi \sigma^3 N/(6L^3)$. A
natural time scale is the rotational relaxation time
$\tau_r=1/(2D_r)$. Despite its simplicity, the model captures the
main aspects of various real systems, such as thermo- or
diffusiophoretic spherical microswimmers
\cite{howse2007prl,jiang2010prl,theurkauff2012prl,buttinoni2013prl}
with $Pe\approx1-80$, persistently swimming spherical colonies 
of the green algae {\it Volvox} \cite{drescher2009prl} with $Pe\approx10^7$ 
and spherical bacteria {\it Serratia marcescens} \cite{rabani2013plos} with $Pe\approx10$.

\section{Phase behaviour}
We first focus on a system with large P\'eclet number, $Pe=272$.
This system exhibit a remarkable phase behavior as function of the
global packing fraction $\phi$. Above a critical packing fraction
$\phi_c \approx 0.3375$, the system separates into liquid-like and
gas-like domains. The morphology of the domains is shown
exemplarily in Fig.~\ref{f:isodensity} for different values of
$\phi$. With increasing volume fraction, the morphology varies
from a liquid droplet (Fig.~\ref{f:isodensity}(a)) to a
bicontinuous structure (Fig.~\ref{f:isodensity}(b)) and finally to
a gas bubble (Fig.~\ref{f:isodensity}(c)).

\begin{figure}[t]
\includegraphics[width=1.0\columnwidth]{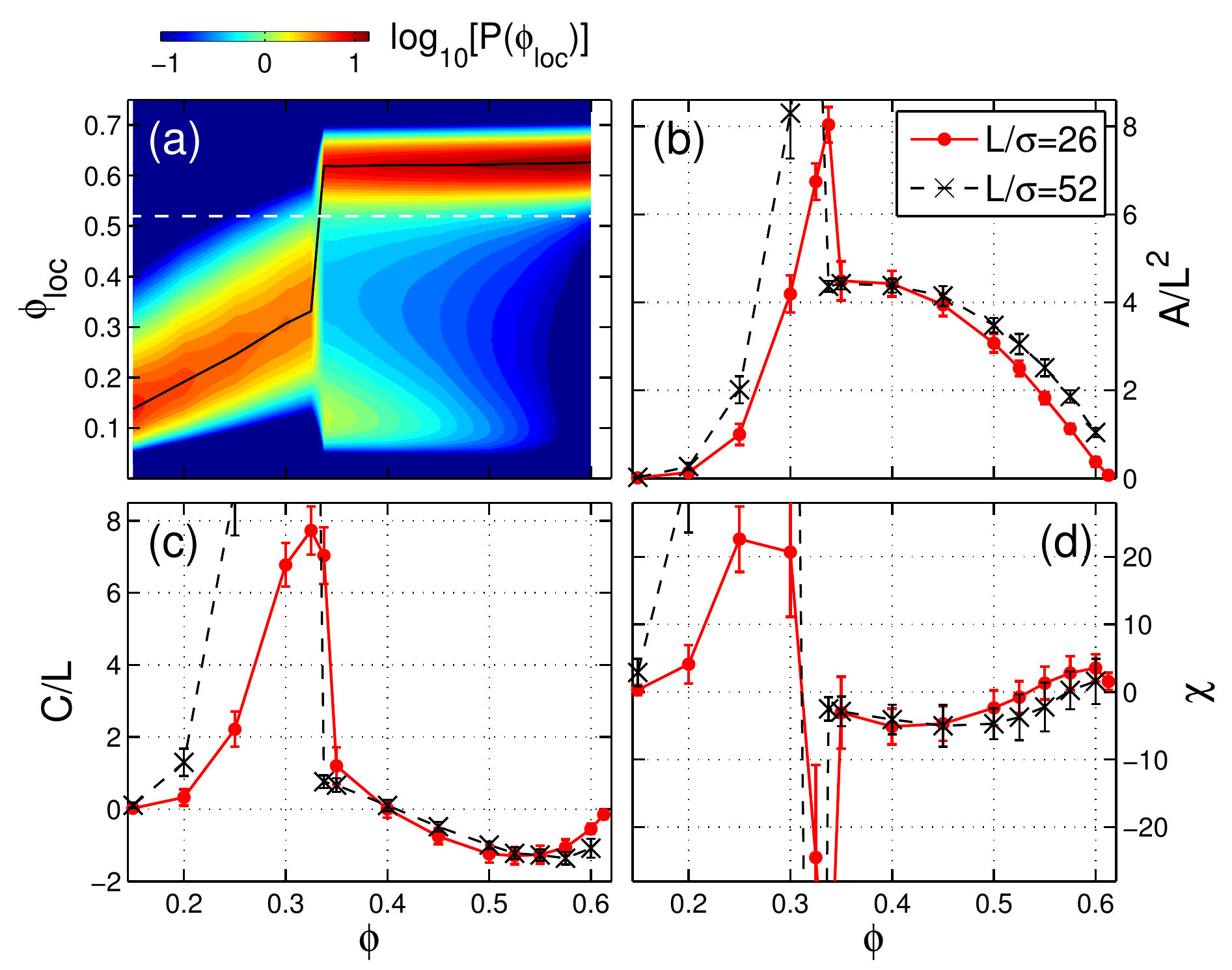}
\caption{\label{f:pdf_pack_minko}
(color online) 
(a) Probability distribution $P(\phi_{loc})$ of the local packing fraction
$\phi_{loc}$ as a function of the global packing fraction $\phi$ for $Pe=272$
and $L/\sigma=52$. The most probable $\phi_{loc}$ is indicated by a solid
black line. The density $\phi_{loc}=0.52$ used as a threshold in 
Fig.~\ref{f:isodensity} is indicated by a dashed line.
(b-d) Morphological analysis of the liquid phase in terms of the
Minkowski functionals interface area $A$ (b),
integrated mean curvature $C$ (c),
and Euler characteristic $\chi$ (d).}
\end{figure}

Figure~\ref{f:pdf_pack_minko}(a) shows the probability
distribution $P(\phi_{loc})$ of the local packing fraction $\phi_{loc}$
as a function of $\phi$, where $\phi_{loc}$ is obtained by Voronoi
construction \cite{rycroft2009chaos}. For $\phi \lesssim \phi_c
\approx 0.3375$, the system is essentially homogeneous and
$P(\phi_{loc})$ is unimodal. While approaching $\phi_c$ with
increasing $\phi$, $P(\phi_{loc})$ broadens (by formation of
transient clusters) and becomes bimodal above $\phi_c$.
Remarkably, the local packing fraction in the liquid phase
$\phi_{liq} \approx 0.62$ is very high, in particular, it is deep within
the glassy region
($\phi_G\approx0.58\leq\phi_{liq}\leq\phi_{RCP}\approx0.64$) of
passive hard spheres \cite{brambilla2009prl,kamien2007prl}. 
Nevertheless, particles remain mobile and no long-range
crystalline order is detected.

The morphological transformations can be characterized by the
surface area $A$, the integrated mean curvature $C$, and the Euler
characteristic $\chi$ of the liquid-gas interface. These
quantities naturally arise in a geometric characterization,
because they are the three (additive) Minkowski functionals in 3D;
for details see Ref.~\cite{michielsen2001pr}. In the two phase
regime, we classify particles as part of the liquid phase if
$\phi_{loc}>0.52$, as indicated by the dashed line in
Fig.~\ref{f:pdf_pack_minko}(a). The average values of all
functionals, as well as their fluctuations, rise dramatically with
increasing $\phi\rightarrow\phi_c$ due to enhanced formation of
transient spherical ($\chi>0$) clusters, see
Fig.~\ref{f:pdf_pack_minko}(b-d). At $\phi_c$, the clusters start
to form a percolating network and consequently $\chi < 0$.
With increasing $\phi>\phi_c$, the liquid phase forms first a
nearly bicontinous structure ($\chi<0$ and $C<0$), whose surface
is reminiscent of the Schwarz P surface, and finally at high
$\phi$, gas-phase droplets float in a dense matrix ($\chi>0$ and
$C<0$). In contrast to an equilibrium system (away from the
critical point) the morphology of the gas (or liquid) phase is
highly dynamic as is reflected by the large fluctuations of
$\chi$, i.e., the structures continuously merge and break as
illustrated by the movies {\tt movie1.avi} and {\tt movie2.avi}.

The structure formation of motile hard sphere particles in 3D
displays an essential difference compared to 2D: clusters in 2D are crystalline \cite{redner2013prl}
at sufficiently high $Pe$, while clusters in 3D are fluid.
Note that rapidly compressed hard spheres form a glass phase \cite{davis1989science},
which is absent in 2D for monodisperse discs.
This has important consequences for the structure and the dynamics. In both 2D and 3D,
particles at the rim of the cluster point inward
and hence exert pressure on the interior. However, in the 2D case, the interior cannot easily react to
an inhomogeneous pressure distribution because of its finite elastic moduli.
This is different for a fluid interior in 3D;
a inhomogeneous pressure generates internal currents and a continuous deformation
of the structure.
\begin{figure}[t]
\centering
\includegraphics[width=0.7\columnwidth]{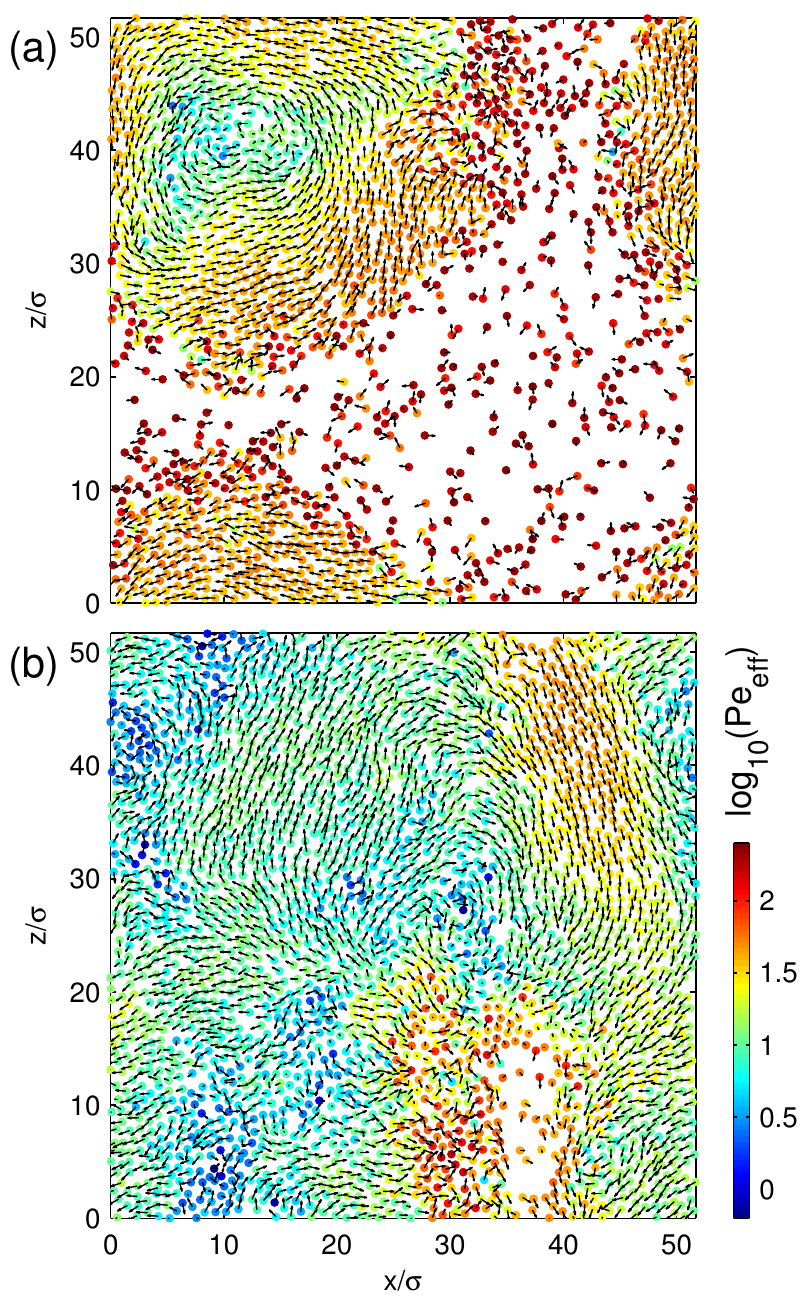}
\caption{\label{f:snapshots}(color online) Collective motion in the steady state at $Pe=272$,
see supplementary movies {\tt movie3.avi} and {\tt movie4.avi}.
Snapshots (slices of thickness $\sigma$) of a system (a) just above the clustering transition ($\phi=0.3375$)
and (b) at a high concentration ($\phi=0.6$).
Arrows indicate the direction of the displacements
$\mathbf{d}(t,\Delta t)=\mathbf{r}(t+\Delta t)-\mathbf{r}(t)$ over a
lag time $\Delta t\approx0.4\tau_r$. The magnitude $d=|\mathbf{d}|$ is color-coded
and is expressed as an effective P\'eclet number, $Pe_{eff}=d/(\Delta t\sigma D_r)$.}
\end{figure}

\section{Collective dynamics}
The phase-separated systems reveal an intriguing dynamics within
the liquid phase (cf. Fig.~\ref{f:snapshots}). Large-scale
coherent displacement patterns emerge, with amplitudes and
directions strongly varying spatially. In addition, transient
swirl-like and jet-like structures appear frequently.
Figure~\ref{f:snapshots}(a) shows a swirl spanning the whole
cluster while Fig.~\ref{f:snapshots}(b) displays a large mobile
region moving between the gas-phase regions of the system, see
supplementary movies {\tt movie3.avi} and {\tt movie4.avi}. This
is surprising considering the fact that
$\phi_{liq}\approx0.62>\phi_G\approx0.58$.
\begin{figure}[t]
\includegraphics[width=1.0\columnwidth]{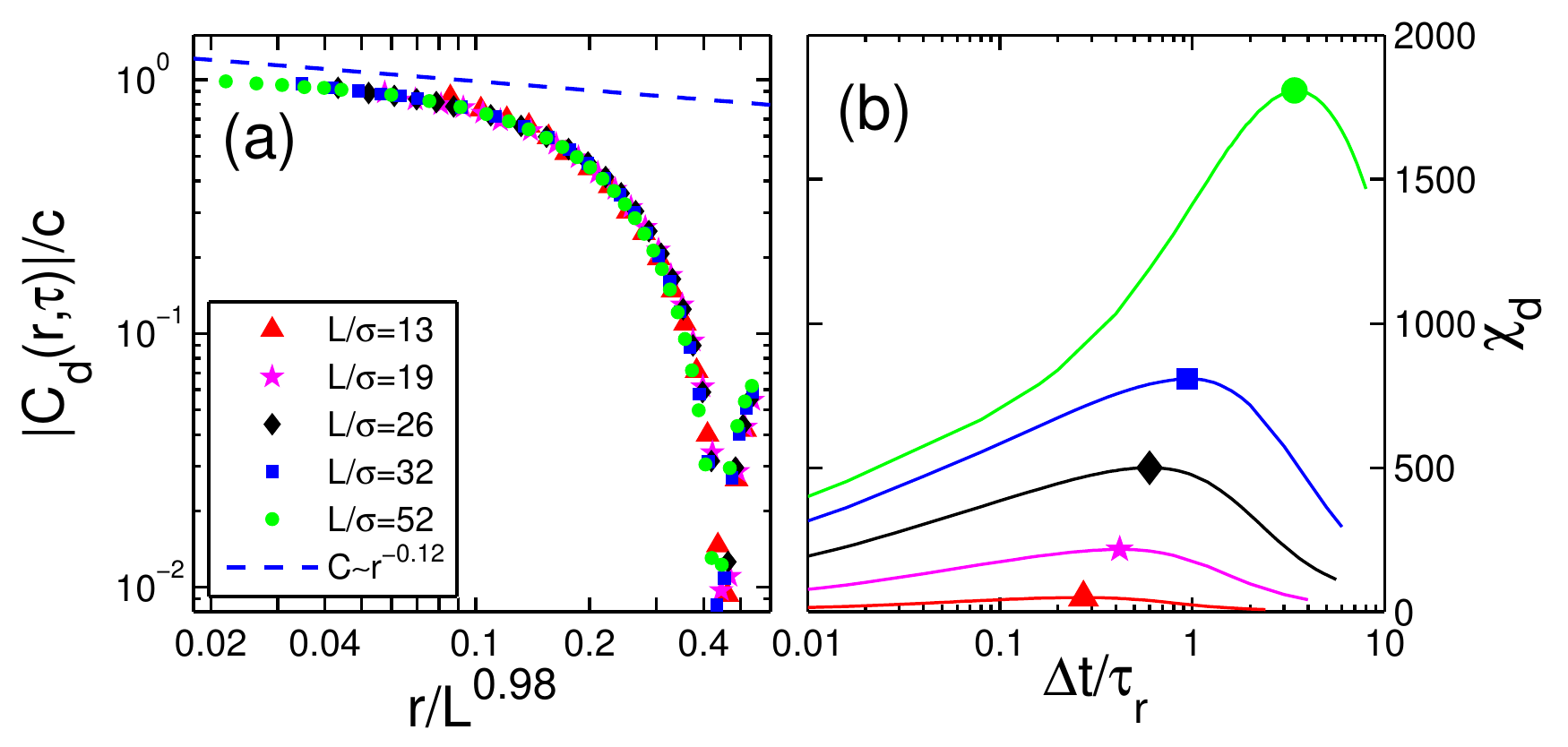}
\caption{\label{f:Cd_rt_Chi_t}(color online) (a) Modulus of the spatial displacement correlation function
$C_{d}(r,\Delta t=\tau)$ as a function of $r/L^{0.98}$ for $\phi=0.6$ and $Pe=272$.
We show $C_{d}$ at the time $\tau$ of maximal cooperativity.
$C_d$ is scaled by $c=0.46,0.56,0.57,0.52,0.33$ for $L/\sigma=13,19,26,32,52$, respectively.
(b) Cooperativities $\chi_d(\Delta t)$ of the various system sizes $L$.
 The symbols indicate $\tau$.}
\end{figure}

The particle dynamics displays a strong heterogeneity in space and time. To
characterize the correlated dynamics, we adopt the
spatio-temporal correlation function \cite{doliwa2000pre}
\begin{equation}\label{e:correlation}
C_{d}(r,\Delta t)=\frac{\left\langle\sum_{i,j\ne i}
\mathbf{d}_i\cdot\mathbf{d}_j \,
\delta(r-\left|\mathbf{r}_{i}-\mathbf{r}_{j}\right|)\right\rangle_t}
{c_0\left\langle\sum_{i,j\ne i}\,
\delta(r-\left|\mathbf{r}_{i}-\mathbf{r}_{j}\right|)\right\rangle_t},
\end{equation}
where $\mathbf{d}_i(t,\Delta t)=\mathbf{r}_i(t+\Delta
t)-\mathbf{r}_i(t)$ is the displacement of particle $i$ over a lag
time $\Delta t$ and $c_0=\left\langle
\sum_i\mathbf{d}_i^2/N\right\rangle_t$ is a normalization factor.
Due to the finite system size we measure $\mathbf{d}$ in the
center-of-mass reference frame. In the limit $\Delta t
\to 0$, $C_d(r,\Delta t)$ becomes the equal-time
spatial velocity correlation function. Moreover, we introduce a
global measure of cooperative motion as 
\begin{equation} \label{e:cooperativity}
\chi_{d}(\Delta t)=\frac{N}{L^3}\int_{\sigma}^{L/2} 4\pi r^2C_{d}(r,\Delta t)\,\mathrm{d}r,
\end{equation}
which can be interpreted as the average number of coherently moving neighbors.
The cooperative motion is not instantaneously present, it rather builds up with time.
As shown in Fig.~\ref{f:Cd_rt_Chi_t}(b), $\chi_{d}(\Delta t)$ first increases,
reaches a maximum at $\Delta t = \tau$ (as indicated by symbols), and then
decays again.
The bell-like shape of $\chi_d(\Delta t)$ is characteristic for the dynamics
in dense liquids \cite{doliwa2000pre}. The origin of this behavior is 
that at short times particles move
independently in their own cage (low correlation),
but gradually they feel the motion of their neighbors and
the displacements become correlated over an increasing length scale.
\begin{figure*}[t]
\centering
\includegraphics[width=1.5\columnwidth]{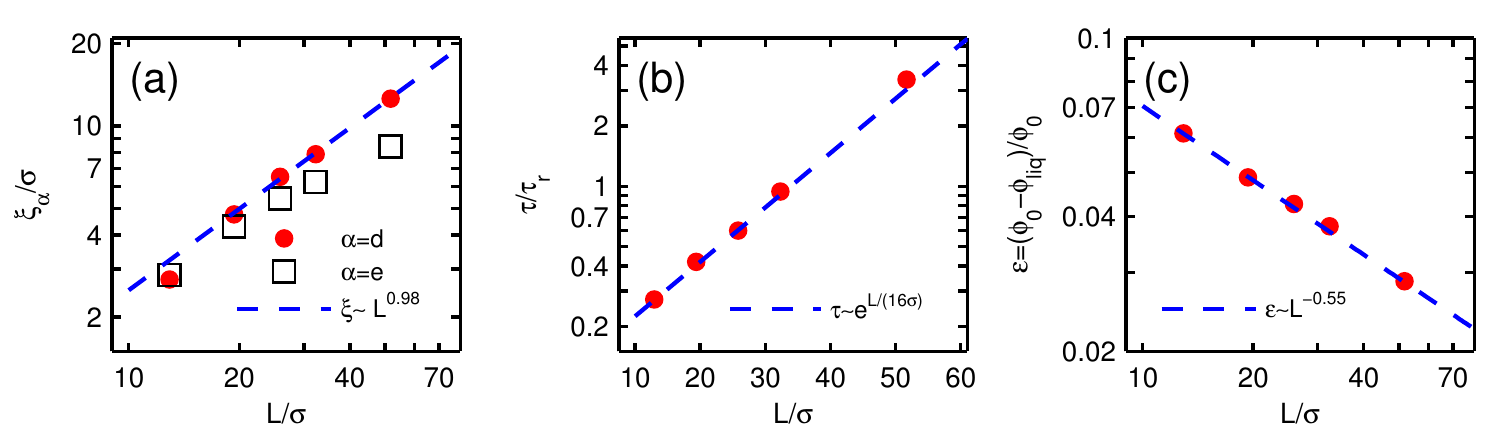}
\caption{\label{f:xi_tau_pack_L}(color online) (a) Correlation
lengths of the displacements $\xi_d(\tau)$ and the orientations
$\xi_e$ as a function of system size $L$. (b) Exponential grow of
the build-up time $\tau$ with $L$. (c) Local density of the liquid
phase $\phi_{liq}$ increases with $L$. The same parameter as in
Fig.~\ref{f:Cd_rt_Chi_t}, $\phi=0.6$ and $Pe=272$.}
\end{figure*}

In the following, we focus exemplarily on a system with
$\phi=0.6$, where on average one or two gas-phase droplets float
within the dense liquid matix ($\phi_{liq} \approx 0.626$), see
Fig.~\ref{f:isodensity}(c) and Fig.~\ref{f:snapshots}(b); yet our
conclusions are valid for most $\phi>\phi_c$.
Figure~\ref{f:Cd_rt_Chi_t}(a) shows $C_{d}$ for various system
sizes $L$ at the time $\tau$ of maximal cooperativity. 
The correlations exhibit an initial power-law decay,
followed by an exponential decay, and become finally negative.
The anticorrelation corresponds to a backflow of particles, which
can be attributed to the incompressibility of the liquid phase.

Fitting a stretched exponential function (along with a power law),  
\begin{equation}
C(r) \sim r^{-\alpha} \exp{\left[-(r/\xi)^\beta\right]},
\end{equation}
to the data yields the exponents $\alpha=0.12$ and $\beta=1.3$, and 
a (nearly) linear increase of the correlation length $\xi_{d}(\tau)$ 
with $L$ (cf.~Fig.~\ref{f:xi_tau_pack_L}(a)).
The good data collapse in Fig.~\ref{f:Cd_rt_Chi_t}(a) with scaled
distance $r/L^{0.98}$ indicates the existence of an universal 
finite-size scaling function, at least over the considered range of 
system sizes.  Furthermore, for large $L$, the system is expected 
to become scale invariant, with a power-law decay of $C_d(r)$. 
A similar behavior has been discussed in other active systems
\cite{cavagna2010pnas,chen2012prl},
however, with a notable polar alignment mechanisms, such as
starling flocks \cite{cavagna2010pnas} and motile bacteria
colonies \cite{chen2012prl}.

Active fluids with repulsive interactions between particles
resemble in some aspects passive fluids in equilibrium with
attractive interactions, e.g., both display a phase
separation \cite{tailleur2008prl,cates2013epl,stenhammar2013prl,stenhammar2014sm,speck2013arxiv}.
However, in many other aspects, active fluids behave very
differently. First, we find an increase of the local density of
the liquid phase $\phi_{liq}$ with system size $L$ at fixed $\phi$ as
$(\phi_0-\phi_{liq})/\phi_0\sim L^{-1/\delta}$ with $\delta\approx1.8$
and $\phi_0\approx0.645$ (note that the actual $\phi$ is somewhat
larger due to the Yukawa part of the potential), see
Fig.~\ref{f:xi_tau_pack_L}(c).
Such a {\em self-compaction} is a clear non-equilibrium feature of
active fluids. Second, we observe a strong increase of the build-up time
$\tau$ with increasing $L$, namely, $\tau\sim\exp{(L/\xi_0)}$ with
$\xi_0=16\sigma$,
see Fig.~\ref{f:xi_tau_pack_L}(b). Hence, long-living, coherently moving
domains are expected for large systems. In fact, these two
observations can be related to each other by the physics of dense
passive fluids near the glass transition.
It is generally accepted that the relaxation time (proportional the build-up time) near the glass
transition obeys the generalized Vogel-Fulcher-Tammann law,
$\tau\sim\exp{[A(\phi_0/(\phi_0-\phi_{liq}))^\delta]}$ with $\delta=1-2$
and $\phi_0=0.614-0.637$ \cite{brambilla2009prl}. However, particle activity
fluidizes the
glass phase and shifts the glass transition to $\phi_0=0.6573$
with $\delta=1$ in the limit $Pe\rightarrow\infty$ \cite{ni2013nc}.
Thus, the large correlation times $\tau$ for large systems can
be understood as a consequence of (active) self-compaction.

Intuitively, a strong correlation between the particle orientation
$\mathbf{e}$ and its direction of motion $\mathbf{d}/|\mathbf{d}|$
can be expected. However, we find only a weak relation between the
cooperative motion and orientation correlations. The latter can be
characterized by the (equal-time) correlation function $C_e(r)$,
where the displacement
vector $\mathbf{d}$ is replaced by the orientation vector
$\mathbf{e}$ in Eq.~(\ref{e:correlation}). In addition, a global measure
$\chi_e$ of orientation correlations can be defined in analogy to
$\chi_d$ in Eq.~(\ref{e:cooperativity}). Although, $C_e$ signals
long-range correlations, however, weaker than $C_{d}$, with $\xi_{e}<\xi_{d}$ (see Fig.~\ref{f:xi_tau_pack_L}(a)), the orientation
correlations, which originate from small regions of polar order at
the gas-liquid interface, are very weak and $\chi_e$ 
is an order of magnitude smaller than $\chi_d(\tau)$. 

Motivated by snapshots (Fig.~\ref{f:snapshots}) and movies ({\tt
movie3.avi} and {\tt movie4.avi}), we propose the following
mechanism behind the observed collective dynamics. Consider first
a smooth (flat) interface between the gas and liquid phase: fast
moving gas-phase particles bump into the interface and remain
there until rotational diffusion leads to an approximately
parallel orientation of $\mathbf{e}$ to the interface
\cite{elgeti2013epl}. In concave regions of a curved interface,
however, the reorientation of $\mathbf{e}$ requires more time
until escape is possible, i.e., particles are effectively trapped.
By contrast, particles in convex regions (protrusions) slide
easily into a neighboring concave regions. This self-amplifying
process accumulates particles with distinct polar order in concave
regions and thereby triggers streams of orientationally disordered
particles deep inside the liquid phase, which explains the much weaker correlations in particle 
orientations than in displacements.
Due to mass conservation,
the particles stream into a neighboring gas droplet or toward the
droplet which itself activated the flow. As a result, further
deformations of droplets are produced and hence new instabilities
are induced.

\begin{figure}[t]
\includegraphics[width=1.0\columnwidth]{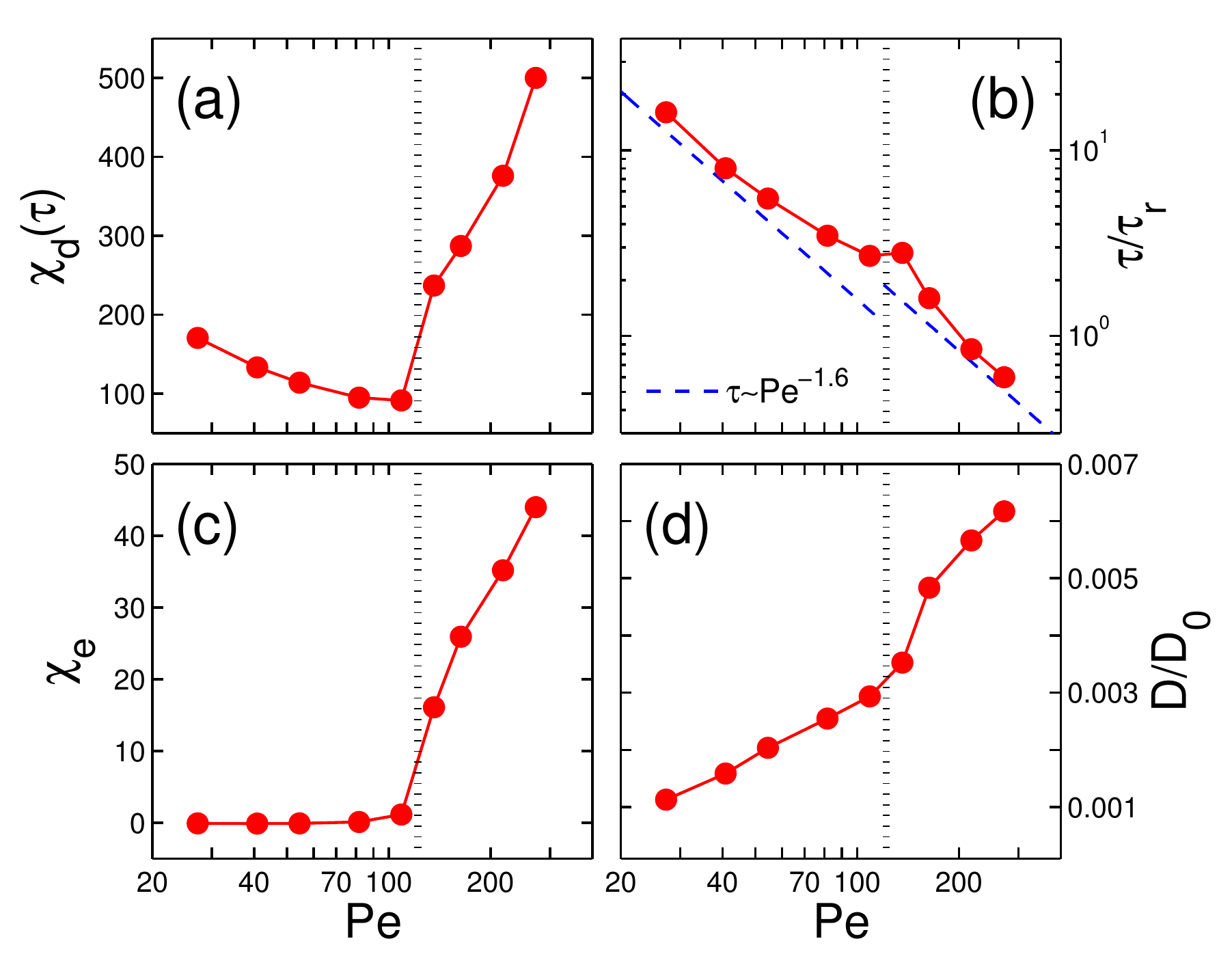}
\caption{\label{f:chi_tau_Pe}(color online) (a) Cooperativity $\chi_{d}(\tau)$,
(b) build-up time $\tau$, (c) global polar correlation $\chi_{e}$ and
(d) long-time diffusion constant $D$ (normalized by $D_0=D_t+V_0^2/(6D_r)$)
as a function of $Pe$ for the packing fraction $\phi=0.6$ and the system size $L/\sigma=26$.
The dashed line indicates $Pe_{c,1}\approx 120$.}
\end{figure}

\begin{figure}[t]
\centering
\includegraphics[width=0.8\columnwidth]{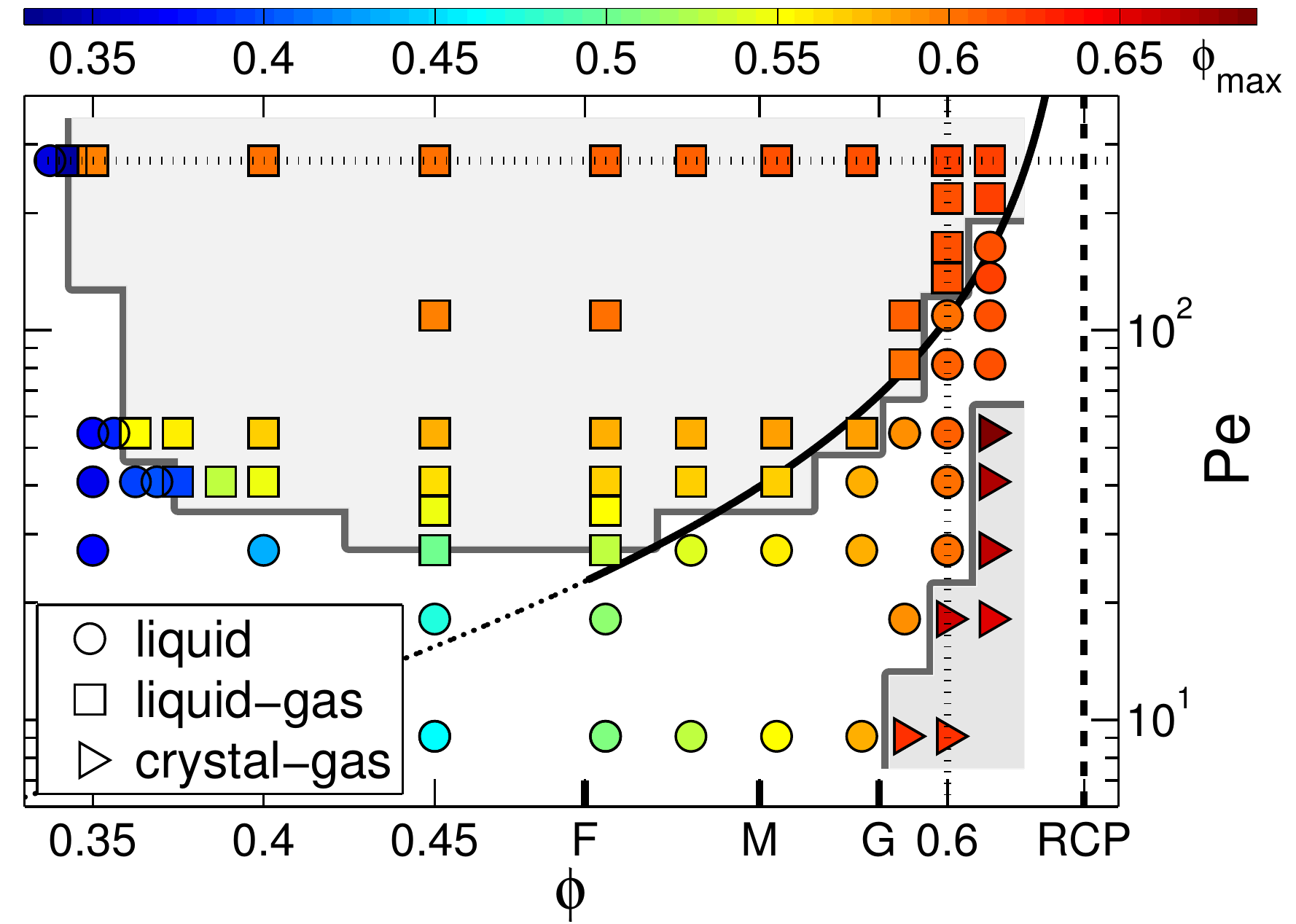}
\caption{\label{f:phase_diag}(color online) Phase diagram of the active suspension spanned by $Pe$ and $\phi$.
Symbols denote the homogenous liquid phase ($\circ$),
the gas-liquid coexistence ($\Box$) and the crystal-gas coexistence 
($\triangleright$).
The equilibrium transition points of hard-sphere freezing ($\phi_F=0.494$), melting ($\phi_M=0.545$),
glass-transition point ($\phi_G\approx0.58$), and random close packing ($\phi_{RCP}\approx0.64$), are marked
by F, M, G, RCP, respectively. The most probable $\phi_{loc}$ is color-coded, i.e.,
in the two-phase region one can read off the density of the dense liquid phase $\phi_{liq}$.
The horizontal and the vertical short dashed lines correspond to results presented
in Fig.~\ref{f:pdf_pack_minko} and Fig.~\ref{f:chi_tau_Pe}, respectively.
The solid line marks $Pe_c(\phi)$ which is proportional to Eq.~(\ref{e:pressure}).}
\end{figure}

So far we have focussed on the large P\'eclet number $Pe=272$.
However, variation of the activity substantially affects the
structure and dynamics of the suspension. We illustrate this
behavior for $\phi=0.6$ (see vertical dashed line in
Fig.~\ref{f:phase_diag}). By decreasing $Pe$ we pass three
regions, a liquid-gas phase at high $Pe$, a homogeneous
liquid phase below $Pe_{c,1}\approx 120$, and finally a
crystal-gas phase below $Pe_{c,2}\approx 20$. In the homogeneous
phase $Pe_{c,2}<Pe<Pe_{c,1}$, there is no polar order and
$\chi_{e}\approx 0$, see Fig.~\ref{f:chi_tau_Pe}(c). The dynamical
behavior of the systems change drastically while crossing
$Pe_{c,1}$. As displayed in Fig.~\ref{f:chi_tau_Pe}(a), the
maximal cooperativity $\chi_{d}(\tau)$ is nearly constant below
$Pe_{c,1}$ and increases strongly with $Pe$ above $Pe_{c,1}$. In
addition, activity substantially speeds up the build-up time
$\tau$, namely, $\tau\sim Pe^{-1.6}$ as shown in
Fig.~\ref{f:chi_tau_Pe}(b). As a result, collective motion below
$Pe_{c,1}$ appears on time scales $\tau \gg \tau_r$ and becomes
increasingly reminiscent of an passive supercooled fluid
\cite{doliwa2000pre}; note that $\tau=\infty$ for $Pe=0$ since
$\phi>\phi_G$ \cite{brambilla2009prl}. Finally, the particle
dynamics is drastically slowed down for $Pe< Pe_{c,2}$, where the
system starts to freeze, as indicated in the phase diagram in
Fig.~\ref{f:phase_diag}. In Fig.~\ref{f:chi_tau_Pe}(d), we show
the diffusivity $D$ normalized by the corresponding value of a
free active particle, $D_0=D_t+V_0^2/(6D_r)$. Although the system
is above $\phi_G$, $D$ is finite and is an increasing function of
$Pe$. The latter suggests a shift of the glass transition to
larger $\phi$ with increasing activity
\cite{kranz2010prl,berthier2013np,ni2013nc,berthier2013arxiv}.

We collect all our results in a phase diagram spanned by $Pe$ and
$\phi$, see Fig.~\ref{f:phase_diag}. For high $Pe$, the homogenous
phase is unstable over a broad range of packing fractions $\phi$,
as explained above. The unstable range decreases with decreasing
$Pe$ (the lower bound is predicted to obey $\phi_c \sim \phi_1+\phi_2/Pe$
\cite{fily2013sm,redner2013prl} with constants $\phi_{1,2}$).
Below $Pe_{min}\approx 30$ phase separation disappears for all $\phi$.
Except for large $\phi$ and
low $Pe$, where crystalline domains are formed, the dynamics is
fluid-like.  The surprising feature of the phase diagram at high
$\phi$, the presence of a homogeneous fluid phase in the gap between the 
liquid-gas and the crystal-gas coexistence, was very recently also
observed in an active soft-disk system in 2D \cite{fily2013sm}.
Finally, we expect jamming to occur for $\phi \to \phi_{RCP}$ in the
limit of $Pe \to \infty$.

The phase separation as a function of $\phi$ and $Pe$ can be
understood as follows. The pressure of a hard
sphere fluid increases with $\phi$ and diverges at RCP for the
metastable branch as
\begin{equation}\label{e:pressure}
p_{HS}=-\frac{6k_BT}{\pi\sigma^3}\phi^2\frac{\mathrm{d}}{\mathrm{d}\phi}\ln{\left\{\left[\left(\phi_{RCP}/\phi\right)^{1/3}-1\right]^3\right\}},
\end{equation}
according to the free volume theory \cite{kamien2007prl}.
Self-propelled particles at low $\phi$ easily overcome this
pressure, coagulate due to their slow-down during collisions
(overdamped dynamics) and hence form clusters. The density within
the cluster, $\phi_{liq}$, adjusts such that $p_{HS}(\phi_{liq})$ balances
the active pressure $p_a$. An initially homogenous system can only
phase separate if the active pressure
$p_a\sim\gamma_tV_0/\sigma^2$ exceeds $p_{HS}(\phi)$, this
explains the existence of a boundary, $Pe_c(\phi)$, between the liquid-gas coexistence and the homogenous liquid phase at high $\phi$, 
see solid line in Fig.~\ref{f:phase_diag}; a similar argument was used
in Ref.~\cite{fily2013sm} in the context of soft disks.

\section{Summary and Conclusions}
Suspension of active particles without an alignment rule exhibit
on broad range of fascinating structural and collective dynamical
effects. In particular, we find a phase separation in a gas phase
and a liquid phase with a highly dynamic domain structure. The
interior dynamics of the liquid domains displays a strong
heterogeneity in space and time. Surprisingly, the dynamical
behavior is highly collective, despite the absence of an alignment
mechanism. This collective dynamics can only arise from a spatial
sorting of particles with similar orientation. The reorientation
of particles at an interface provides such a sorting mechanism.

Several interesting questions arise from our results. In
particular, the crossover from the equilibrium phase behavior to
the regime at low $Pe$ deserves further attention.
The mechanism of self-compaction, which does not seem to be
explainable easily from neither an effective active particle attraction
nor from the compressive interface layer, has to be elucidated. Studies of
active fluids at low $Pe$ may also shed light on the dynamics in glassy
systems in a complementary way to external fields like shear.

\acknowledgments
We thank M. Abkenar, T. Auth, S. Henkes, J. Horbach, R. Ni, S. Poblete,
G. M. Sch\"utz and M. Sperl for helpful discussions.
Financial support by the VW Foundation (VolkswagenStiftung) within the program
{\it Computer Simulation of Molecular and Cellular Bio-Systems as well as Complex Soft Matter} is gratefully acknowledged.

\bibliographystyle{eplbib}

\end{document}